\documentclass{article}

\usepackage{microtype}
\usepackage{graphicx}
\usepackage{subfigure}
\usepackage{booktabs} % for professional tables
\usepackage{bm}
\usepackage{hyperref}

\usepackage[accepted]{icml2025}

\usepackage{amsmath}
\usepackage{amssymb}
\usepackage{mathtools}
\usepackage{amsthm}
\usepackage{bbm}

\usepackage[capitalize,noabbrev]{cleveref}

\theoremstyle{plain}

\theoremstyle{definition}

\theoremstyle{remark}

\renewcommand{\vec}[1]{\ensuremath{\bm{#1}}}

\newcommand{\mat}[1]{\ensuremath{\mathbf{#1}}}

\usepackage[textsize=tiny]{todonotes}

\icmltitlerunning{MoE-CE: Enhancing Generalization for Deep Learning based Channel Estimation  via a Mixture-of-Experts Framework}

\begin{document}

\twocolumn[
\icmltitle{MoE-CE: Enhancing Generalization for Deep Learning based \\ Channel Estimation via a Mixture-of-Experts Framework}

% It is OKAY to include author information, even for blind
% submissions: the style file will automatically remove it for you
% unless you've provided the [accepted] option to the icml2025
% package.

% List of affiliations: The first argument should be a (short)
% identifier you will use later to specify author affiliations
% Academic affiliations should list Department, University, City, Region, Country
% Industry affiliations should list Company, City, Region, Country

% You can specify symbols, otherwise they are numbered in order.
% Ideally, you should not use this facility. Affiliations will be numbered
% in order of appearance and this is the preferred way.
% \icmlsetsymbol{equal}{*}

\begin{icmlauthorlist}
\icmlauthor{Tianyu Li}{comp}
\icmlauthor{Yan Xin}{comp}
\icmlauthor{Jianzhong (Charlie) Zhang}{comp}
\end{icmlauthorlist}

\icmlaffiliation{comp}{Standards and Mobility Innovation Lab, Samsung Research America, Berkeley Heights, New Jersey, USA}

\icmlcorrespondingauthor{Tianyu Li}{tianyu.li@partner.samsung.com}

% You may provide any keywords that you
% find helpful for describing your paper; these are used to populate
% the "keywords" metadata in the PDF but will not be shown in the document
\icmlkeywords{Machine Learning, ICML}

\vskip 0.3in
]

% this must go after the closing bracket ] following \twocolumn[ ...

% This command actually creates the footnote in the first column
% listing the affiliations and the copyright notice.
% The command takes one argument, which is text to display at the start of the footnote.
% The \icmlEqualContribution command is standard text for equal contribution.
% Remove it (just {}) if you do not need this facility.

\printAffiliationsAndNotice{}  % leave blank if no need to mention equal contribution
% \printAffiliationsAndNotice{\icmlEqualContribution} % otherwise use the standard text.

\begin{abstract}
Reliable channel estimation (CE) is fundamental for robust communication in dynamic wireless environments, where models must generalize across varying conditions such as signal-to-noise ratios (SNRs), the number of resource blocks (RBs), and channel profiles. Traditional deep learning (DL)-based methods struggle to generalize effectively across such diverse settings, particularly under multitask and zero-shot scenarios. In this work, we propose MoE-CE, a flexible mixture-of-experts (MoE) framework designed to enhance the generalization capability of DL-based CE methods. MoE-CE provides an appropriate inductive bias by leveraging multiple expert subnetworks, each specialized in distinct channel characteristics, and a learned router that dynamically selects the most relevant experts per input. This architecture enhances model capacity and adaptability without a proportional rise in computational cost while being  agnostic to the choice of the backbone model and the learning algorithm. Through extensive experiments on synthetic datasets generated under diverse SNRs, RB numbers, and channel profiles, including multitask and zero-shot evaluations, we demonstrate that MoE-CE consistently outperforms conventional DL approaches, achieving significant performance gains while maintaining efficiency.
\end{abstract}

\section{Introduction}
In modern wireless communication systems, reliable data transmission over time-varying and multi-path fading channels depends critically on 
the accurate knowledge of the channel state information (CSI). Channel estimation (CE) plays a vital role in this context by enabling the receiver to mitigate distortions introduced by the wireless medium. CE techniques are particularly important in advanced communication systems such as the fifth generation (5G) cellular networks, massive multiple input multiple output (MIMO), and millimeter-wave communications, where maintaining high data rates, low latency, and spectral efficiency is essential despite complex propagation conditions. Traditional channel estimation methods, including least squares (LS) and minimum mean square error (MMSE)~\cite{neumann2018learning}, rely heavily on pilot symbols and statistics of the channel models. However, the growing complexity of modern wireless environments has motivated the exploration of data-driven and learning-based approaches that can capture intricate channel behaviors and adapt to dynamic conditions more effectively. Nevertheless, these models often struggle to generalize across diverse deployment scenarios, such as varying signal-to-noise ratio (SNR) levels, the number of resource blocks (RBs), or channel profiles, particularly when task distribution shifts significantly at inference time.

\begin{figure}[t]
\vskip 0.2in
\begin{center}
\centerline{\includegraphics[width=\columnwidth]{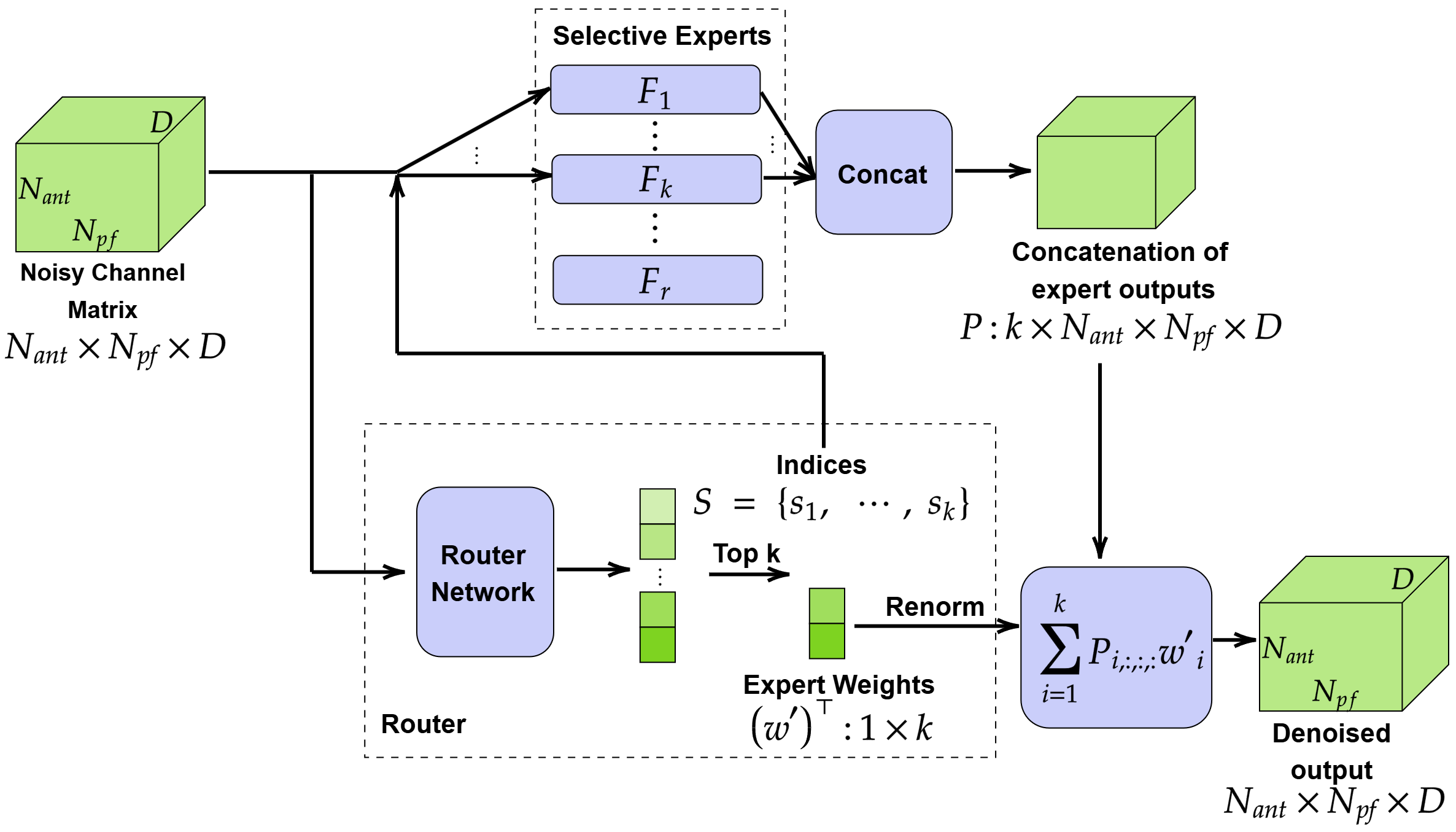}}
\caption{MoE framework for channel estimation.}
\label{fig:moe_ce_pipeline}
\end{center}
\vskip -0.2in
\end{figure}

Several efforts have been made to enhance the generalization capabilities of deep learning (DL)-based CE approaches. From a data-centric standpoint,~\cite{luan2023achieving} highlights the critical role of training dataset design in achieving robust performance across varying channel conditions. Their work shows that exposing DL models to a rich diversity of simulated channel environments during training significantly improves their adaptability to unseen deployment scenarios. From a learning algorithm perspective, in~\cite{mao2019roemnet}, the authors introduce a meta-learning-based approach designed to improve channel estimation in orthogonal frequency division multiplexing (OFDM) systems called RoemNet. It employs a meta-learner that can adapt to varying channel conditions, demonstrating superior performance compared to traditional methods under diverse scenarios. Despite these promising directions, many existing approaches emphasize data and training strategies while neglecting the role of model architecture itself. Architectural inductive biases, such as modularity, or conditional computation, are rarely explored as a primary avenue for improving generalization. Designing models that inherently support robustness to channel variability, even under standard training regimes, remains an open and underexplored challenge in the literature.

Mixture-of-experts (MoE) is a neural network architecture that introduces conditional computation by dynamically selecting a subset of specialized sub-models, or ``experts'', for each input. Originally proposed in~\cite{jacobs1991adaptive}, the core idea is rooted in the divide-and-conquer principle, where a complex problem is decomposed into smaller, more manageable subproblems. 
%This structure enables the model to allocate computational modules and resources adaptively and learn diverse representations tailored to different input patterns. 
Unlike monolithic models that must use a shared set of weights for all data variations, MoE architectures encourage modularity, allowing different experts to specialize in distinct subspaces of the task distribution. 
This architectural inductive bias not only improves efficiency but also enhances the model’s ability to generalize across heterogeneous tasks and input domains.
In multitask setting, MoE has demonstrated strong performance by capturing task-specific structure while maintaining flexibility, as shown in domains such as robotics robotics~\cite{huang2025moe} and computer vision~\cite{chen2023adamv}.
This architectural flexibility makes MoE particularly well-suited for zero-shot generalization as well, where the model must handle tasks or channel conditions that were not explicitly seen during training. The routing mechanism enables input-dependent expert selection, allowing the model to adaptively decompose complex learning problems and respond flexibly to new or unseen scenarios, as demonstrated in~\cite{muqeeth2024learning}.
This property is especially valuable in domains like wireless channel estimation, where rapid adaptation to changing environments is critical and explicit supervision for every possible condition is impractical.

In this paper, we propose mixture-of-experts framework for channel estimation (MoE-CE), an MoE framework designed to enhance the generalization capability of DL-based channel estimation methods. The MoE-CE architecture comprises multiple expert sub-networks, each specializing in different channel characteristics or task variations, alongside a gating mechanism that dynamically routes input features to the most relevant experts. This modular design not only expands model capacity without a proportional increase in computational complexity but also enables flexible adaptation to a wide range of channel conditions.

Importantly, MoE-CE is agnostic to the specific learning algorithm and backbone network. In other words, it can incorporate any deep learning architecture for channel estimation and be trained using a variety of optimization strategies, from standard gradient-based methods with different optimizers to more advanced schemes like meta-learning (such as MAML~\cite{finn2017model}), to further improve the model’s generalization capabilities.

We demonstrate that MoE-CE achieves substantial performance gains in both multitask learning settings, where the model learns across multiple channel types or system configurations; and in zero-shot scenarios, where it generalizes to previously unseen channel conditions at test time. Our results show that the combination of expert specialization and dynamic routing in MoE-CE is particularly well-suited for the non-stationary and diverse nature of wireless channels.

\textbf{Contributions.} The key contributions of this work are outlined as follows:
\begin{itemize}
    \item We introduce MoE-CE, a flexible and learning-algorithm-agnostic framework that enhances the generalization performance of DL-based channel estimation. It supports arbitrary backbone architectures and can be integrated with various learning strategies, including meta-learning, to further boost generalization. 
    \item We evaluate MoE-CE under both multitask and zero-shot settings, showing consistent improvements over conventional DL-based channel estimation (DL-CE) baselines under similar computational complexity.
    \item Using a mixed-SNR training scenario as a case study, we analyze expert selection patterns and demonstrate how MoE-CE promotes specialization and adaptability across diverse channel conditions.
\end{itemize}

\section{Background and Related Work}
To contextualize our proposed framework, this section reviews the foundational concepts and existing literature relevant to our work. We begin by presenting the mathematical formulation of the channel estimation problem in OFDM systems and briefly review recent DL-based approaches developed for this task. Next we introduce the MoE architecture, highlighting its core principles, practical applications in large-scale models, and relevance to channel estimation. Finally, we discuss strategies for managing expert load balancing, including both auxiliary-loss-based and auxiliary-loss-free approaches, which are critical for efficient and stable training of MoE-based systems.

\subsection{Channel Estimation}
OFDM is a widely used modulation technique in modern communication systems due to its robustness against frequency-selective fading and efficient implementation via the Fast Fourier Transform (FFT). In this section, we describe the mathematical formulation of the channel estimation problem in a typical OFDM system.

Consider an OFDM system with $N_{pf}$ pilot sub-carriers and $N_{ant}$ receive antennas. The input-output relationship between transmitted and received signals at
pilot sub-carriers in the frequency domain can be written as:
\begin{equation}
	\mathbf{Y}=\mathbf{H}\odot \mathbf{X}+\mathbf{W},\label{input-output1}
\end{equation}
where $\mathbf{Y}\in \mathbb{C}^{N_{ant}\times N_{pf}}$ are the received signals at $N_{pf}$ pilot sub-carriers and $N_{ant}$ receive antennas, $\mathbf{H}\in \mathbb{C}^{N_{ant}\times N_{pf}}$ denotes the channel matrix in the space-frequency domain, the operator $\odot$ represents the Hadamard product that is an element-wise product,  $\mathbf{X}\in \mathbb{C}^{N_{ant}\times N_{pf}}$ are the transmitted pilot signals known to the receiver, and $\mathbf{W}\in \mathbb{C}^{N_{ant}\times N_{pf}}$ is an additive white Gaussian noise (AWGN).

% Consider an OFDM system with $N_{pf}$ subcarriers and $N_{ant}$ antennas. Let $\textbf{X} \in \mathbb{C}^{N_{pf}\times N_{ant}}$ denote the frequency-domain transmitted signal and denote the channel matrix $\textbf{H}\in \mathbb{C}^{N_{pf}\times N_{ant}}$ and the additive white Gaussian noise (AWGN) $\textbf{W}\in \mathbb{C}^{N_{pf}\times N_{ant}}$ drawn from the distribution $\mathcal{N}(0, \sigma^2\mat{I})$. Then the frequency-domain received signal $\textbf{Y} \in \mathbb{C}^{N_{pf}\times N_{ant}}$ is computed by the following: 
% \begin{equation}
%     \textbf{Y} = \textbf{H}\odot\textbf{X} + \mat{W},
% \end{equation}
% where $\odot$ denotes the Hadamard product.

The goal of the channel estimation task is to estimate channel matrix $\textbf{H}$ based on the pilot signal $\textbf{X}$ and the received signals  $\textbf{Y}$. The simplest channel estimation solution is the LS estimate, denoted by $\hat{\textbf{H}}^{LS}$, which is readily computed by:
\begin{equation}
    \hat{\textbf{H}}^{LS}_{i, j} = \frac{\textbf{Y}_{i, j}}{\textbf{X}_{i, j}},~~\forall i\in[N_{ant}],~j\in[N_{pf}], %,
\end{equation}
where the notation $[N]$ denotes all positive integers no larger than $N$. The above equation can therefore be further simplified to:
\begin{equation}
\label{eq:ls}
    \hat{\textbf{H}}^{LS} = \textbf{H} + \textbf{W}.
\end{equation}

\subsection{DL-based Channel Estimation}
In recent years, DL has emerged as a powerful tool in the field of wireless communications, offering promising enhancements to traditional signal processing algorithms. Various DL-based techniques have been explored to improve tasks such as modulation classification~\cite{8054694}, signal detection~\cite{erdogmus2001nonlinear}, channel equalization~\cite{he2018deep}, and CSI feedback compression~\cite{samuel2017deep}. In the context of channel estimation, conventional estimators like LS and MMSE rely on statistical assumptions and predefined models, which may not generalize well to real-world environments. In contrast, DL-based methods can learn complex mappings directly from data, enabling more flexible and robust estimation.
DL-based channel estimation techniques can be broadly divided into two categories. The first category adopts an end-to-end learning perspective, treating the entire communication system as a differentiable model. For instance, in~\cite{ye2017power}, a deep neural network is trained to perform encoding, decoding, channel estimation and all other functionalities of a
communication link jointly in an implicit fashion, showing significant performance gains over traditional baselines. Similarly,~\cite{8054694} proposes an autoencoder-based communication system that integrates modulation, channel estimation, and decoding into a single trainable model. However, such approaches often lack explicit access to the estimated channel state, thereby limiting their applicability in systems where CSI is needed for other signal processing tasks.
The second category focuses specifically on learning the channel matrix using supervised deep neural networks. In~\cite{wen2018deep}, the authors treat the channel matrix as a two-dimensional (2D) image and apply a convolutional neural network (CNN)~\cite{lecun1998gradient} for denoising-based channel estimation in massive MIMO systems. This method captures spatial correlations across antennas effectively. Following this idea, many recent channel estimation works have adopted this framework, such as~\cite{8944280, soltani2019deep, ahmad2023image, 9288911, 8752012}, proving the importance and effective of this DL framework for channel estimation. 

From the~\cref{eq:ls}, the CE problem can be viewed as a 2D denoising problem in the frequency domain, where the goal is to recover the true channel matrix from its noisy LS estimate. Typically, a neural network is trained to learn a nonlinear mapping from the noisy LS estimate, obtained using pilot symbols, to the underlying clean channel matrix.
\begin{equation}
    \hat{\mathbf{H}} = f_\theta(\hat{\mathbf{H}}^{LS}),
\end{equation}
where $f_\theta$ is a neural network parameterized by $\theta$. 

By decomposing the complex channel matrix into its real and imaginary components, we obtain a tensor of size $N_{ant}\times N_{pf}\times 2$, effectively transforming the channel estimation task into an image denoising problem. Neural networks developed for vision tasks, such as CNNs~\cite{lecun1998gradient} and more advanced architectures like Resnet~\cite{he2016deep}, and NAFNet~\cite{chen2022simple}, are well-suited for this setting, as they can effectively capture spatial correlations across antenna and subcarrier dimensions. When applied to channel estimation, these models are typically trained using the normalized mean squared error (NMSE) as the loss function:
\begin{equation}
    \mathcal{L}_{\text{NMSE}} = \mathbb{E} \left[ \frac{\|\mathbf{H} - \hat{\mathbf{H}}\|_2^2}{\|\mathbf{H}\|_2^2} \right].
\end{equation}

% In recent years, data-driven methods using deep learning have gained attention for their ability to learn complex mappings from pilot observations to channel estimates. CNNs have been applied to exploit local structure in pilot patterns and capture spatial correlations in the channel \cite{ye2017power, chen2022compressive}. Recurrent architectures and transformer-based models have also been explored for time-varying channels and sequential estimation tasks [need reference]. These models often outperform traditional estimators in specific scenarios and can be trained end-to-end. 
% However, they typically lack robustness to domain shifts and exhibit poor generalization when tested on channel conditions that differ significantly from the training distribution.

%To address these limitations, some works have explored transfer learning, meta-learning, and domain adaptation techniques in wireless communication tasks. While these methods improve flexibility, they still rely on either task-specific fine-tuning or shared representations that may not capture the full diversity of real-world channels.
\subsection{Mixture of Experts Architecture}

Modern MoE architectures are often based on transformers and consist of two main elements: sparse MoE layers and a gating network or the so-called router. On one hand, the sparse MoE layer consists of various “expert” blocks. Typically, these expert blocks are parameterized by feed forward networks (FFN) in the transformer architecture, often with the same structure. The router, on the other hand, determines which input is sent to which expert. The router does so by outputting a vector of weights for all the experts. 
A subset of the top-k experts is then selected based on these weights, and only these experts are activated to process the input. Finally, the outputs of the selected experts are aggregated and passed to the next layer.
Notably, the MoE layer often replaces the traditional FFN layer in the Transformer architectures, enabling more specialized modeling without incurring too much computational overhead.
% Then a selection of top $k$ experts is made based on these weights and only these selected experts get to process the input. Finally, the selected experts aggregate their results together and output to the next layer. Note that the MoE layer often replaces the traditional FFN layer in the transformer to allow for more specialized modelling without incurring too much computation overhead. 
Through expert specialization, the FFN in the MoE layer likely requires fewer parameters, 
thereby further enhancing the computational efficiency of the transformer.
% hence it can even further improve the computational efficiency of the transformer. 

In large-scale language models, MoE has been widely adopted to enhance scalability while maintaining computational efficiency. Models such as Switch Transformer~\cite{fedus2022switch}, GShard~\cite{lepikhingshard}, and DeepSeek~\cite{liu2024deepseek} utilize MoE architectures to increase the number of parameters without a proportional rise in computational cost. In these architectures, only a small fraction of the experts are active per token, significantly reducing the per-step float operations (FLOPs) consumption compared to a dense model of similar size. The key benefit of MoE in language models is its ability to scale efficiently while mitigating inference costs. By activating only a subset of experts, MoE architecture enables models to learn diverse representations across different tokens, capturing nuanced patterns in natural language.

% A size $r$ MoE layer with top $k$ selection is defined as: given an input $\vec{x}$, a set of $r$ experts $\{F_1, F_2, \dots, F_r\}$ produce candidate outputs, and a gating function $R(\vec{x}) \in \mathbb{R}^r$ assigns routing probabilities:
% \begin{equation}
%     \hat{\vec{y}} = \sum_{i=1}^r r(\vec{x})_i \cdot F_i(\vec{x}),
% \end{equation}
% where $R(\vec{x})_i \geq 0$ and $\sum_i R(\vec{x})_i = 1$. In hard routing, only the top $k$ experts are activated, reducing computational cost:
% \begin{equation}
%     \hat{\vec{y}} = \sum_{i \in \mathcal{T}_k(\vec{x})} R(\vec{x})_i \cdot F_i(\vec{x}),
% \end{equation}
% where $\mathcal{T}_k(\vec{x})$ denotes the top $k$ experts selected by the router.

An MoE layer with $r$ experts is defined as follows: given an input $\vec{x}$, a set of expert functions $\{F_1, F_2, \dots, F_r\}$ each computing a candidate output, and a gating function $R(\vec{x}) \in \mathbb{R}^r$ assigning routing weights, the output of a fully routed MoE layer is computed as:
\begin{equation}
    \hat{\vec{y}} = \sum_{i=1}^r R(\vec{x})_i \cdot F_i(\vec{x}),
\end{equation}
where $R(\vec{x})_i \geq 0$ and $\sum_{i=1}^r R(\vec{x})_i = 1$. 

In the case of hard routing with top $k$ selection, only the $k$ experts with the highest gating scores are activated, reducing computational cost. The output becomes:
\begin{equation}
    \hat{\vec{y}} = \sum_{i \in \mathcal{T}_k(\vec{x})} R(\vec{x})_i \cdot F_i(\vec{x}),
\end{equation}
where $\mathcal{T}_k(\vec{x})$ denotes the indices of the top $k$ experts selected by the gating function.

\subsection{Load Balancing in MoE}

Training MoE models requires careful handling of load balancing to prevent uneven expert utilization. When certain experts receive a disproportionate amount of traffic, the model's training efficiency and convergence may deteriorate. To address this, prior works have employed auxiliary losses to encourage balanced expert utilization. Notably, GShard~\cite{lepikhingshard} and Switch Transformer~\cite{fedus2022switch} incorporate auxiliary losses that penalize imbalanced expert activation. For instance, GShard utilizes a load balancing loss that discourages excessive reliance on a small subset of experts. Additionally, Switch Transformer simplifies the gating mechanism to a top $1$ selection, reducing the routing overhead while improving load distribution. Specifically, Switch Transformer utilizes the auxiliary loss:
\begin{equation}
\label{eq:switch}
    \mathcal{L}_{\text{load}} = \sum_{i=1}^r\frac{\alpha\cdot N}{T^2}\sum_{\vec{x}\in\mathcal{B}}\mathbbm{1}\{\mathrm{argmax}~R(\vec{x})=i\}\sum_{\vec{x}\in\mathcal{B}}R(\vec{x})_i,
\end{equation}
where $\mathcal{B}$ denote the current batch, $T$ is the number of tokens, $N$ is the number of examples and $\alpha$ is a regularization weight.~\cref{eq:switch} encourages uniform routing since it is minimized under a uniform distribution. 

\subsection{Auxiliary-Loss-Free Load Balancing}

To eliminate the need for explicitly tuning load balancing losses, recent studies have proposed architectural strategies that naturally promote balanced expert usage. One such approach, auxiliary loss-free load balancing (ALFLB), is introduced by DeepSeek~\cite{guo2025deepseek, liu2024deepseek, wang2024auxiliary}. Instead of enforcing load balancing through auxiliary objectives in classic literature, this approach leverages an adaptive gating function that naturally distributes computation across experts. To achieve this, the model needs to maintain an expert bias, initialized as an all-zero vector of size $r$. After each gradient update, one needs to compute the frequency of the expert selection. If the frequency for an expert being selected is higher than a threshold $\tau_1$, we call this expert being “over-utilized”. Otherwise, if it is lower than a threshold $\tau_2$, we refer to it as being “under-utilized”. For the over-utilized experts, we decrease the corresponding expert bias by $\gamma$ while for the under-utilized ones, we increase their expert bias by $\gamma$. When selecting the top $k$ experts, we use the sum of the router’s output weights and the expert bias to determine which experts to select.

\section{Methodology}
In this section, we present the mixture-of-experts framework for
channel estimation (MoE-CE) in details. This framework is flexible and can accommodate any backbone ML models and learning methods. The core idea is to construct multiple expert networks, which can adopt various neural network architectures tailored to different aspects of CE, e.g. different SNR levels, RB numbers, or channel profiles. A lightweight router network is responsible for dynamically selecting the top $k$ most relevant experts during each forward pass, enabling efficient resource allocation and adaptive learning. By dynamically selecting the most suitable experts based on the current input, the model achieves improved generalization and computational efficiency, enhancing the robustness and adaptability of CE models in dynamic communication environments. We will show later in the experiment sections that this framework works well not only under a multitask set-up but also has significant performance gain when testing in a zero-shot setting as well.

\begin{algorithm}[tb]
   \caption{Training MoE-CE with ALFLB and SGD}
   \label{alg:moe-ce}
\begin{algorithmic}[1]
   \STATE {\bfseries Experts (subnetworks):} NN parameterized functions $F_1, \cdots, F_r: \mathbb{R}^{N_{ant}\times N_{pf} \times D}\rightarrow\mathbb{R}^{N_{ant}\times N_{pf} \times D}$.
   \STATE {\bfseries Router:} NN parameterized function $R: \mathbb{R}^{N_{ant}\times N_{pf} \times D}\rightarrow \mathbb{R}^r$.
   \STATE {\bfseries Input:} Noisy channel (LS estimate of the channle matrix) $\hat{\mat{H}}^{LS}\in \mathbb{R}^{N_{ant}\times N_{pf} \times D}$.
   \STATE {\bfseries Input:} Clean channel $\mat{H}\in \mathbb{R}^{N_{ant}\times N_{pf} \times D}$.
   \REPEAT
   \STATE Initialize the expert bias $\vec{u} \in \mathbb{R}_+^r$ for ALFLB to be all zeros.
   \STATE Initialize the router $R$ and the subnetworks $F_1, \cdots, F_r$. 
   \STATE Compute the forward pass of the router function: $\vec{w}^\top = R(\hat{\mat{H}}^{LS})$. 
   \STATE Compute the top $k$ selection weights $\tilde{\vec{w}}^\top = \vec{w}^\top + \vec{u}^\top$
   \STATE Obtain the top $k$ indices based on $\tilde{\vec{w}}$, e.g., $S = \{1, \cdots, k\}$ and the corresponding router weights weights $\vec{w}^{\prime\top}$, e.g., $\vec{w}^{\prime\top} = [\vec{w}_1, \cdots, \vec{w}_k]^\top$.
   \STATE Based on the expert selection frequency, adjust the expert bias $\vec{u}$. 
   \STATE Renormalize the weights: $\vec{w}^{\prime\top} = \frac{\vec{w}^{\prime\top}}{<\vec{w}^{\prime\top}, \textbf{1}>}$. 
   \STATE Compute the forward pass of the selected subnetworks, e.g., $F_{1}(\hat{\mat{H}}^{LS}), \cdots, F_{k}(\hat{\mat{H}}^{LS})$. Stack these candidate outputs to form a tensor $\mat{P}\in \mathbb{R}^{k\times N_{ant}\times N_{pf} \times D}$. 
   \STATE Obtain the final output: $$\sum_{i=1}^k \mat{P}_{i,:,:,:}\vec{w}^\prime_i.$$
   \STATE Compute the NMSE loss function $\mathcal{L}_{\mathrm{NMSE}}$ w.r.t. $\mat{H}$ and run stochastic gradient descent to update the parameters of the selected subnetworks $F_{1}, \cdots, F_{k}$ and the router $R$. 
   \UNTIL{Converge}
\end{algorithmic}
\end{algorithm}

\cref{fig:moe_ce_pipeline} illustrates the MoE-CE pipeline. An MoE-CE top $k/r$ with $r$ selective experts and top $k$ selection goes as the following: first we take as input the LS estimate of the channel matrix either in the frequency domain or the delay domain, which is of size $N_{ant}\times N_{pf} \times D$. Note that typically $D$ is set to $2$ indicating the real and imaginary decomposition of the complex channel matrix. Alternatively, the parameter $D$ can also be $4$ when polarization is introduced. The input will first go through a router network $R: \mathbb{R}^{N_{ant}\times N_{pf} \times D}\rightarrow \mathbb{R}^r$, mapping the input data to a size $r$ vector. Then after applying softmax function, we obtain the expert weights $\vec{w}^\top$ and subsequently select the top $k$ experts based on the corresponding expert weights and obtain the selected expert indices $S = \{s_1, \cdots, s_k\}$ as well as their corresponding expert weights $\vec{w}^{\prime\top}\in \mathbb{R}^k$. For the ease of illustration, let us assume $s_i = i$, for all $i\in[k]$, i.e., $S = \{1, 2,\cdots, k\}$ . Note that the expert selection varies based on the current input. After obtaining the selected expert indices, the LS estimate input will go through the forward pass of the selected expert networks, namely $F_1, \cdots, F_r:\mathbb{R}^{N_{ant}\times N_{pf} \times D}\rightarrow\mathbb{R}^{N_{ant}\times N_{pf} \times D}$, which can be parameterized as arbitrary neural networks. Assuming $S = \{1, 2, \cdots, k\}$, the input only goes through the selected $k$ expert networks, i.e., $F_1, \cdots, F_k$ and for the remaining $r-k$ networks, no forward pass is carried out. We then obtain $k$ candidate outputs of size $\mathbb{R}^{N_{ant}\times N_{pf} \times D}$ and concatenate them together to form a tensor $\mat{P}\in \mathbb{R}^{k\times N_{ant}\times N_{pf} \times D}$. The selected expert weights $\vec{w}^{\prime\top}$ are renormalized $\vec{w}^{\prime\top} = \frac{\vec{w}^{\prime\top}}{<\vec{w}^{\prime\top}, \textbf{1}>}$, where $\textbf{1}$ is an all $1$ vector of size $k$, so that it sums up to $1$. Finally, a weighted sum is computed to obtain the final denoised channel $\sum_{i=1}^k\mat{P}_{i,:,:,:}\vec{w}^\prime_i$. For resolving load balancing issue, we opt to use ALFLB as introduced in the previous section. We further explain the complete training procedure of MoE-CE with ALFLB and stochastic gradient descent (SGD) in~\cref{alg:moe-ce}. We defer the integration of MoE-CE with alternative learning schemes, such as meta-learning, for future work. 

\begin{figure}[t]
\vskip 0.2in
\begin{center}
\centerline{\includegraphics[width=\columnwidth]{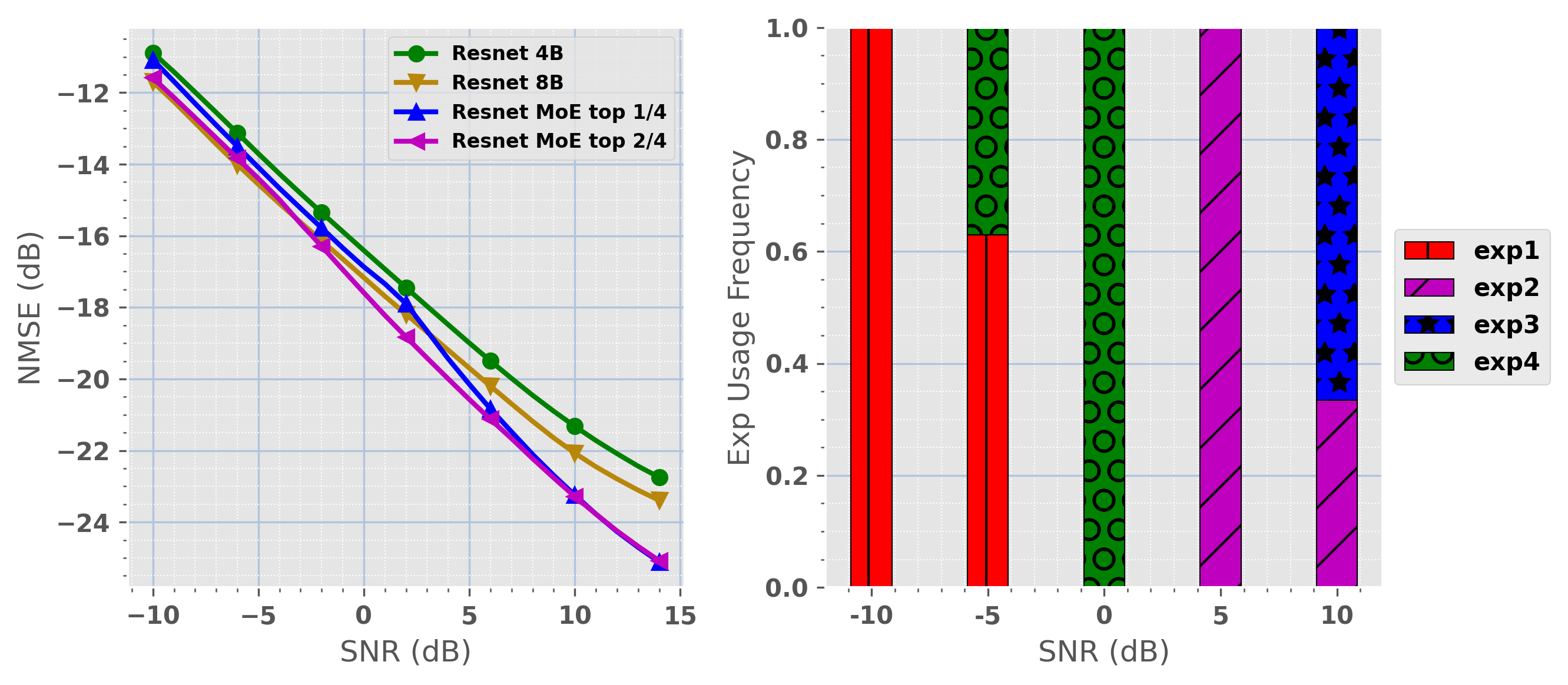}}
\caption{Cross SNR levels channel estimation performance comparison between vanilla Resnet and Resnet-MoE~(left) and expert usage analysis (right).}
\label{fig:re_msnr}
\end{center}
\vskip -0.4in
\end{figure}

\begin{figure*}[t]
\vskip 0.2in
\begin{center}
\centerline{\includegraphics[width=0.9\textwidth]{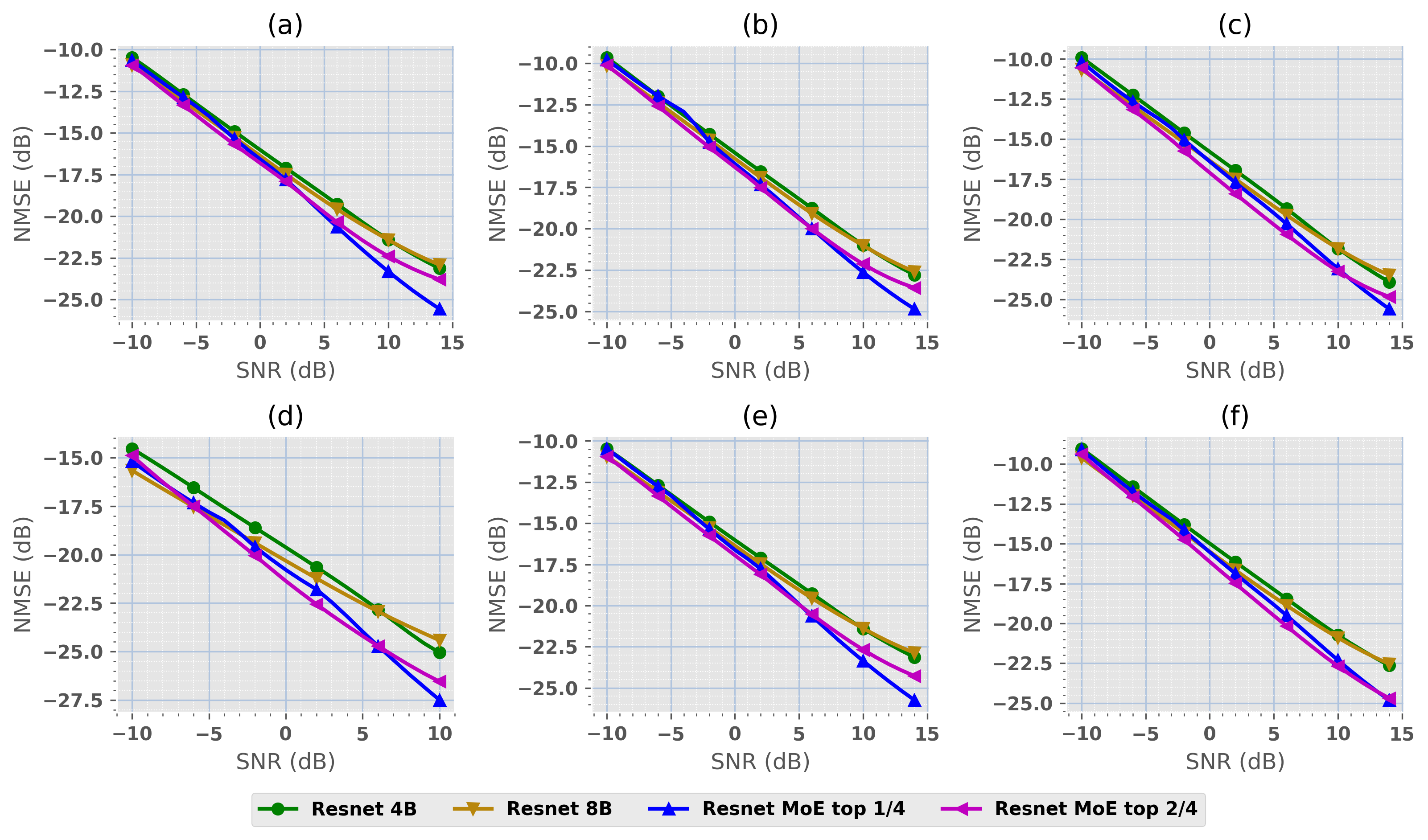}}
\caption{Cross channel profiles channel estimation performance comparison between vanilla Resnet and Resnet-MoE under multitask setting on: (a) UMi, (b) UMa, (c) CDL-B and (d) CDL-D; and under zero-shot generalization setting on: (e) CDL-B with $1200$ ns delay spread and (f) UMi with $1200$ delay spread.}
\label{fig:re_cp}
\end{center}
\vskip -0.28in
\end{figure*}

\section{Experiment}
Generalizing to multiple SNR levels, channel profiles, and RB numbers is vital for building robust and scalable channel estimation models in modern wireless communication systems. Real-world environments are dynamic, with fluctuating SNR and varying channel characteristics due to mobility, interference, and deployment scenarios. A model that can generalize across these variations ensures consistent performance without retraining, reducing latency, and improving system reliability. Likewise, generalization across RB numbers allows the model to adapt to different bandwidth allocations, enabling flexibility and efficiency across standards like 5G and future 6G systems. Such generalization not only lowers deployment complexity and cost by avoiding the need for specialized models but also supports long-term adaptability to evolving network conditions, spectrum usage, and hardware configurations.

\begin{table*}[t]
\caption{Model computational complexity and size comparison between MoE-CE and vanilla DL methods. The complexity is computed based on an input size of $16\times240\times4$ (with polarization).}
\label{tab:complexity}
\vskip 0.15in
\begin{center}
\begin{small}
\begin{sc}
\begin{tabular}{lcccr}
\toprule
Model & Macs & FLOPs & \#Parameters & Model Size \\
\midrule
Resnet MoE top $1/4$    &$79.43$ M&$158.86$ M&$81.19$ K&$317$ KB \\
Resnet MoE top $2/4$ &$159.28$ M&$318.56$ M&$81.19$ K&$317$ KB\\
% Resnet MoE top 1/8    &79.43M&158.86M&164.14k&641kb \\
% Resnet MoE top 2/8 &159.28M&318.56M&164.14k&641kb\\
Resnet $4$B   &$78.84$ M&$157.68$ M&$19.99$ K&$78$ KB \\
Resnet $8$B   &$155.03$ M&$310.06$ M&$38.80$ K&$151$ KB  \\
\hline
NAFNet MoE top $1/4$    &$22.2$ M&$44.4$ M&$147.78$ K&$577$ KB\\
NAFNet MoE top $2/4$    &$44.81$ M&$89.62$ M&$147.78$ K&$577$ KB \\
NAFNet Vanilla     &$21.61$ M&$43.22$ M&$36.64$ K&$145$ KB   \\
\bottomrule
\end{tabular}
\end{sc}
\end{small}
\end{center}
\vskip -0.1in
\end{table*}

When generalizing to different RB numbers or channel profiles, it is often necessary to also generalize to various SNR levels simultaneously because these dimensions of variability are inherently intertwined in real-world wireless environments. For example, a change in RB number affects frequency resolution and spectral efficiency, but its impact on performance is highly dependent on the SNR: what works well at high SNR may fail at low SNR due to significant noise level discrepancy. Similarly, different channel profiles (e.g., urban vs. rural or line-of-sight vs. non-line-of-sight) exhibit distinct multipath and fading characteristics, whose effects are amplified or diminished depending on the SNR. Therefore, to ensure robust channel estimation under realistic deployment scenarios, the model must be capable of handling the joint variability of RB number, channel profile, together with different SNR levels.

In this section, we conduct a series of experiments to evaluate the effectiveness and generalization capability of the proposed MoE-CE framework. We design our evaluations to reflect realistic challenges in wireless communication, including varying signal-to-noise ratios (SNRs), channel profiles, and RB numbers. For experiments involving diverse channel profiles and RB numbers, we also incorporate cross-SNR generalization to simulate more practical and demanding deployment scenarios. The experiments are organized to test both multitask learning performance and zero-shot generalization ability. We compare MoE-CE against strong deep learning baselines under matched computational budgets, and analyze expert utilization patterns to provide insights into the model's adaptability and specialization behavior across diverse conditions.

Throughout the experiment section, we use Adam~\cite{kingma2014adam} with learning rate $0.001$ as the optimizer. The training and test data are generated from physical uplink shared channel (PUSCH) using Siona~\cite{hoydis2022sionna}. For ALFLB setup, we set $\tau_1 = \frac{2}{r}$, $\tau_2 = \frac{4}{5r}$ and $\gamma = 0.001$ via cross validation, where $r$ is the number of selective experts. For all MoE-CE models presented in this section, we use a three-layer CNN as the router architecture, with $3\times 3$ filter size and $r$, $2r$, and $r$ hidden channels in the respective layers. The output of the CNN router is passed through global average pooling followed by a softmax function to produce the initial task weights.

\subsection{Data Preprocessing and Postprocessing}
The data we obtained from the Siona-based system level simulator are the clean channel matrix (the label) and the LS estimate (noisy) of the channel matrix, under the PUSCH frequency domain and OFDM format. We first convert the data into delay domain using fast Fourier transformation (FFT). The ML models then operate on the delay domain transformed data. After obtaining the output from the ML model, an inverse FFT (IFFT) is performed to recover the prediction back to the frequency domain. The loss function as well as the evaluation metric is then computed under the frequency domain.

\subsection{Mixed SNRs}
In this experiment, we showcase the generalization capability of the proposed MoE framework on cross SNRs setups. To be more specific, we generate synthetic training data from urban micro (UMi) channel profile, with SNR ranges from $-10$ dB to $12$ dB, taken every $2$ dB. The RB number for both training and test data is set to be $40$. For evaluation, we look at the NMSE result on a separate test dataset of the same RB number and channel profile, with SNR ranges from $-10$ dB to $14$ dB. For this experiment, we use Resnet~\cite{he2016deep} as the backbone model, namely all experts are parameterized by the Resnet architecture. Specifically, for baseline models, we use Resnet with $4$ Resnet blocks (Resnet-$4$B) and $8$ blocks (Resnet-$8$B), and the channel size is set to $16$. For all MoE experts, we use Resnet-$4$B with $16$ channel size. In the following experiments, without other specifications, all MoE experts are using Resnet-$4$B with $16$ channel size as the architecture. We use $3\times 3$ kernel size for all CNN layers in all experiments.

In the left figure of~\cref{fig:re_msnr}, we show the NMSE comparison of our Resnet-MoE architecture with top $1$ and top $2$ expert selection out of $4$ selective experts. The notation top $1/4$ means we are using top $1$ selection, and the total number of selective experts is $4$. Note that under this set-up, Resnet MoE top $1/4$ and $2/4$ share similar computation complexity to Resnet-$4$B and $8$B, respectively. We can see clearly that MoE based models achieve better performance, especially under relatively high SNR cases. 
% One other thing to observe is that as we increase the selected number of experts, the overall behavior of MoE seems to be improving the performance under low SNR cases while slightly decreasing the performance under high SNR cases. That is another trade-off when one considers implementing an MoE based architecture. 
% The best overall performance is achieved by a top 2/8 architecture, as it strikes a balance between the high SNR cases as well as low SNR cases. 
% This suggests that when there is enough training budget/memory, it is often good to have more experts to choose from, especially note that the computational cost for a forward pass of top 2/4 and top 2/8 are the same. 

Notice that the curve of top $1/4$ (blue curve with upward triangle markers) is not very smooth. This is because different experts are selected at different SNR levels. Due to the discrete nature of the top $1$ selection, we can observe the sharp turning point on this curve. The right figure of Figure~\ref{fig:re_msnr} illustrates the distribution of expert usage during evaluation across different SNR levels. At $-10$ dB SNR, the router chooses expert $1$ for all evaluation input data. This behavior is gradually shifted to expert $4$ at $0$ dB SNR. Then at $5$ dB SNR the router shifts the selection entirely to expert $2$. This is subsequently changed to a mix of expert $2$ and expert $3$ at $10$ dB SNR. Overall, experts $1$ and $4$ are often selected for low SNR cases while experts $2$ and $3$ are for high SNR ones.

In~\cref{tab:complexity}, we present the computational complexity of all models compared in the experiment section. One can clearly see that Resnet MoE top $1/4$ shares similar FLOPs with Resnet $4$B, while Resnet MoE top $2/4$ shares similar FLOPs with Resnet $8$B. One thing to note is that MoE-CE architecture leverages several subnetworks simultaneously, therefore it significantly increases the number of parameters and the model size.

\subsection{Mixed SNRs and Channel Profiles}
% Generalizing to multiple channel profiles in channel estimation is crucial for ensuring robustness to real-world variability, as wireless channels are influenced by environmental factors such as terrain, mobility, and interference. It enables scalability across diverse scenarios, including urban, rural, indoor, outdoor, low- and high-mobility environments, allowing a single model to operate effectively without retraining. Additionally, generalization improves adaptability to dynamic channel conditions caused by user movement, weather changes, and network load, maintaining high performance without frequent updates. It also reduces training and deployment costs by eliminating the need for multiple specialized models. A well-generalized model enhances performance in unknown environments by interpolating or extrapolating to unseen conditions, ensuring reliable operation in unpredictable network deployments. Moreover, it supports next-generation wireless networks, such as 6G, where highly dynamic and heterogeneous environments demand robust models capable of handling different frequency bands, hardware configurations, and propagation conditions.

In this experiment, we want to evaluate the ability of the proposed MoE architecture to generalize to multiple SNR levels and channel profiles, such as UMi, urban macrocellular channel (UMa), clustered delay line (CDL)-B and CDL-D. To do this, we generate synthetic data from all these channel profiles under PUSCH, with SNR ranges from $-10$ dB to $12$ dB. The RB number for both training and evaluation is still set to $40$ and with delay spread set to $300$ ns, $300$ ns, $600$ ns, $10$ ns for the respective channel profiles. For the expert’s architecture, we follow the same Resnet structure as in the previous experiment ($4$ blocks with $16$ channels). For a multitask setting, subfigures (a), (b), (c), and (d) of~\cref{fig:re_cp} present evaluation result of the trained model on UMi, UMa, CDL-B, CDL-D with the same delay spread as the training data respectively. For a zero-shot setting, subfigures (e), (f) show the evaluation results of the same trained model but on CDL-B and UMi with $1200$ ns delay spread instead. We can see that MoE-based approach consistently outperforms the vanilla method, under both multitask setting as well as zero-shot adaptation setting. The gain in performance is more significant with increased SNR, the same phenomenon as we observed in the previous experiment. An additional observation is that increasing the number of selected experts tend to enhance the performance of the MoE model in the low SNR regime, while causing a slight performance degradation in the higher regime. This highlights a trade-off that needs to be carefully considered when implementing an MoE-based architecture.

\subsection{Mixed SNRs and Varying RB Numbers}
% Generalizing to multiple Resource Block (RB) sizes in channel estimation is essential for ensuring flexibility in network deployment, as different wireless systems allocate varying RBs based on bandwidth, user demand, and service requirements. It enhances scalability across wireless standards, such as 5G NR, which dynamically adjust RB allocations to optimize spectrum usage. Additionally, generalization improves adaptability to dynamic bandwidth allocation, where RBs change due to adaptive modulation, traffic balancing, and spectrum efficiency strategies. A single model that generalizes well across RBs reduces computational costs, storage, and deployment complexity, avoiding the need for separate models. Furthermore, since different RBs affect the resolution and frequency selectivity of channel estimation, a generalized model can learn to handle these variations, improving robustness in diverse conditions. Lastly, as future wireless technologies increasingly rely on flexible spectrum allocation, a model capable of generalizing to different RBs ensures long-term compatibility and efficient operation.

\begin{figure}[t]
\vskip 0.25in
\begin{center}
\centerline{\includegraphics[width=\columnwidth]{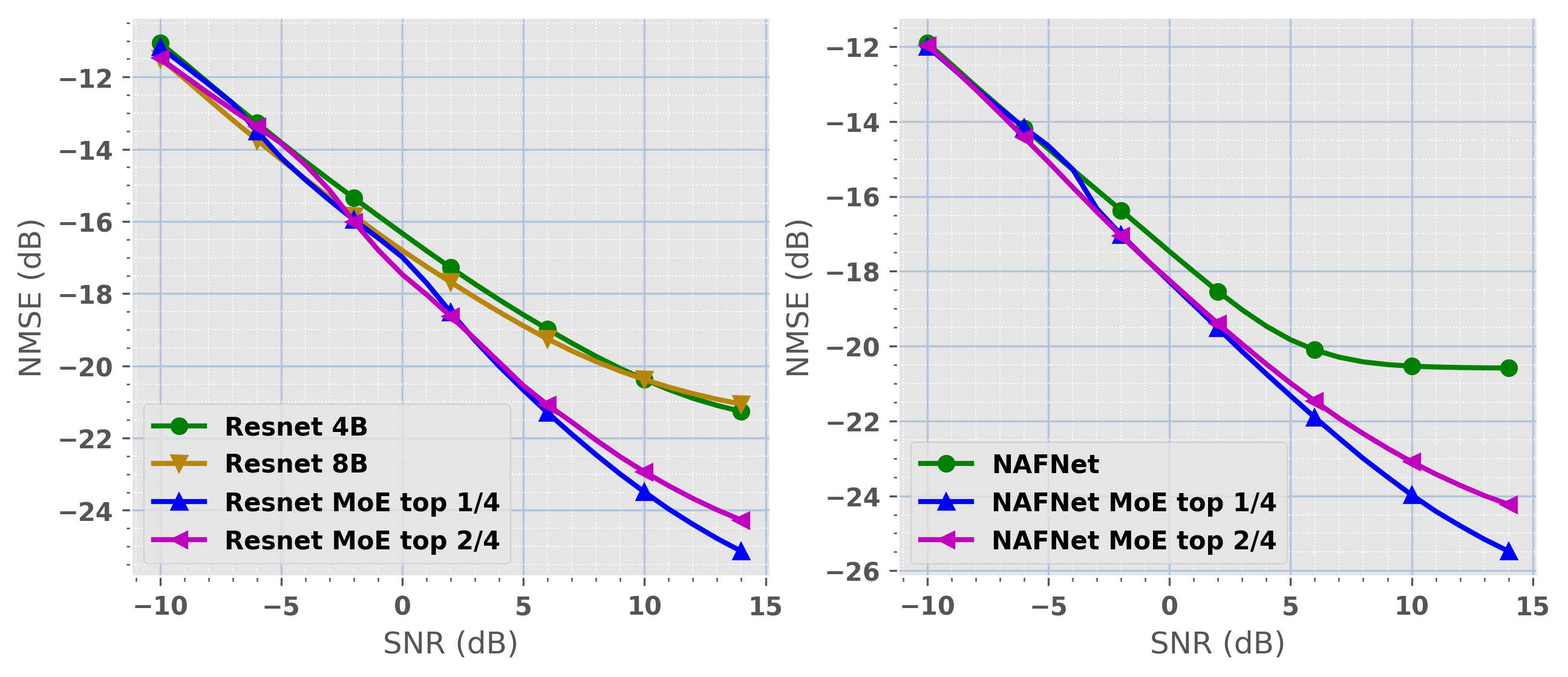}}
\caption{Zero-shot generalization performance comparison for varying RB numbers between vanilla Resnet and Resnet-MoE~(left) and vanilla NAFNet and NAFNet-MoE (right).}
\label{fig:re_mrb}
\end{center}
\vskip -0.3in
\end{figure}

In this experiment, besides showing the generalization capability of the MoE architecture to multiple SNR levels and RB numbers, we also showcase the flexibility of the proposed MoE structure. In addition to using Resnet as the backbone model, we also use a $15$-layer NAFNet~\cite{chen2022simple} with $8$ hidden channels as the backbone model. 
For this experiment, we use the training data generated from a UMi channel with RB numbers of $5, 9, 12, 16, 20$ with SNR ranges from $-10$ dB to $12$ dB. The validation data has the same setup as the training data. 
% We will compare the performance under both delay domain. 
For evaluation, we generate data from the same UMi channel but with $54$ RBs. For Resnet, compared between Resnet $4$B, $8$B and Resnet MoE top $1/4$, $2/4$. Same as before, Resnet MoE utilizes a Resnet $4$B as the backbone model. Additionally, we also compare between vanilla NAFNet and NAFNet MoE of $4$ selective experts with top $1$ and top $2$ selections. Note that this experiment is purely zero-shot setting as the test RB number does not exist in the training RB numbers. 

\cref{fig:re_mrb} presents the cross RB numbers performance evaluation of Resnet backbone (left) and NAFNet backbone (right) for testing the $54$ RB configuration under the delay domain. We can see that in both cases there is a significant performance gain using the MoE-CE architecture compared to the vanilla model. In~\cref{tab:complexity}, we show the computational complexity and model size for NAFNet-MoE. We observe that, same as Resnet-MoE, the forward pass complexity between the vanilla NAFNet and NAFNet-MoE top $1/4$ is nearly identical, albeit the small router complexity. Though the model size and number of parameters is four times the vanilla NAFNet. 

\section{Conclusion}
In this paper, we propose MoE-CE, a mixture-of-experts framework designed to enhance the generalization ability of DL-based channel estimation methods in wireless communication systems. By combining multiple expert networks with a lightweight router for dynamic expert selection, MoE-CE enables task-specific specialization while maintaining computational efficiency. We demonstrated that this architecture generalizes well across a wide range of scenarios, including varying SNR levels, RB numbers, and channel profiles, under both the multitask and zero-shot settings. Extensive experiments showed that MoE-CE consistently outperforms conventional DL-based methods, achieving improved accuracy with comparable computational cost. Furthermore, the framework’s compatibility with diverse backbones and training strategies makes it a flexible and scalable solution for real-world deployment in dynamic and heterogeneous wireless environments.

Although the proposed MoE-CE framework exhibits robust generalization across varying SNR levels, RB numbers, and channel profiles, several promising avenues remain open for future research. 
One immediate possibility is to incorporate model-agnostic meta-learning (MAML)~\cite{finn2017model} or other meta learning algorithms into MoE-CE, further enhance zero-shot generalization capability of the framework. Furthermore, in the current setup, all experts share the same backbone structure. Designing heterogeneous or dynamically configurable expert architectures, e.g., various approaches mentioned in~\cite{han2021dynamic}, could allow for more fine-grained specialization, potentially improving both performance and efficiency. Last but not least, developing formal theoretical understanding of how and why experts specialize, especially under cross-condition generalization, could offer insights on the channel estimation problem under such system configuration, and additionally for more principled design of future MoE-based architectures for communication systems.

\newpage
\section*{Impact Statement}
This work aims to advance the field of machine learning for wireless communication by improving the generalization capability of deep learning-based channel estimation through a modular MoE framework. By enabling robust performance across diverse and dynamic environments, our approach has the potential to enhance the reliability and adaptability of future wireless systems, including 5G and beyond. Improved channel estimation may lead to more efficient use of spectrum and energy, benefiting both infrastructure providers, end users as well as the environment.

We do not anticipate any direct negative societal consequences or ethical concerns from this work. However, as with all technologies that improve the performance of communication systems, care must be taken to ensure equitable access and responsible deployment.

\bibliography{bib}
\bibliographystyle{icml2025}

%%%%%%%%%%%%%%%%%%%%%%%%%%%%%%%%%%%%%%%%%%%%%%%%%%%%%%%%%%%%%%%%%%%%%%%%%%%%%%%
%%%%%%%%%%%%%%%%%%%%%%%%%%%%%%%%%%%%%%%%%%%%%%%%%%%%%%%%%%%%%%%%%%%%%%%%%%%%%%%
% APPENDIX
%%%%%%%%%%%%%%%%%%%%%%%%%%%%%%%%%%%%%%%%%%%%%%%%%%%%%%%%%%%%%%%%%%%%%%%%%%%%%%%
%%%%%%%%%%%%%%%%%%%%%%%%%%%%%%%%%%%%%%%%%%%%%%%%%%%%%%%%%%%%%%%%%%%%%%%%%%%%%%%
% \newpage
% \appendix
% \onecolumn
% \section{You \emph{can} have an appendix here.}

% You can have as much text here as you want. The main body must be at most $8$ pages long.
% For the final version, one more page can be added.
% If you want, you can use an appendix like this one.  

% The $\mathtt{\backslash onecolumn}$ command above can be kept in place if you prefer a one-column appendix, or can be removed if you prefer a two-column appendix.  Apart from this possible change, the style (font size, spacing, margins, page numbering, etc.) should be kept the same as the main body.
%%%%%%%%%%%%%%%%%%%%%%%%%%%%%%%%%%%%%%%%%%%%%%%%%%%%%%%%%%%%%%%%%%%%%%%%%%%%%%%
%%%%%%%%%%%%%%%%%%%%%%%%%%%%%%%%%%%%%%%%%%%%%%%%%%%%%%%%%%%%%%%%%%%%%%%%%%%%%%%

\end{document}